\documentclass[nonacm]{acmart} 

\usepackage{graphicx}
\usepackage{multicol,multirow}
\usepackage{xcolor}
\usepackage{tabularx}
\begin{document}

\title{How public datasets constrain the development of diversity-aware news recommender systems, and what law could do about it.}

\author{Max van Drunen}
\email{m.z.vandrunen@uva.nl}
\orcid{0000-0003-3917-6655}
\affiliation{%
  \institution{Institute for Information Law, University of Amsterdam}
  \city{Amsterdam}
  \country{the Netherlands}
}

\author{Sanne Vrijenhoek}
\email{s.vrijenhoek@uva.nl}
\orcid{0000-0002-1031-4746}
\affiliation{%
  \institution{Institute for Information Law. University of Amsterdam}
  \city{Amsterdam}
  \country{the Netherlands}
}

\renewcommand{\shortauthors}{van Drunen and Vrijenhoek}

% There is no need to include ORCID IDs in your .pdf; this information is captured by the submission portal when a manuscript is submitted. 
%\author[1]{Max van Drunen}
%\author[2]{Sanne Vrijenhoek}

%\authormark{van Drunen \& Vrijenhoek}

%\address[1]{\orgdiv{Institute for Information Law}, \orgname{University of Amsterdam}, \orgaddress{\city{Amsterdam}, \postcode{1018WV},  \country{the Netherlands}}}

%\address[2]{\orgdiv{Institute for Information Law}, \orgname{University of Amsterdam}, \orgaddress{\city{Amsterdam}, \postcode{1018WV},  \country{the Netherlands}}}

%\authormark{van Drunen \& Vrijenhoek}

\begin{abstract}News recommender systems increasingly determine what news individuals see online. Over the past decade, researchers have extensively critiqued recommender systems that prioritise news based on user engagement. To offer an alternative, researchers have analysed how recommender systems could support the media's ability to fulfil its role in democratic society by recommending news based on  editorial values, particularly diversity. However, there continues to be a large gap between normative theory on how news recommender systems should incorporate diversity, and technical literature that designs such systems. We argue that to realise diversity-aware recommender systems in practice, it is crucial to pay attention to the datasets that are needed to train modern news recommenders. We aim to make two main contributions. First, we identify the information a dataset must include to enable the development of the diversity-aware news recommender systems proposed in normative literature. Based on this analysis, we assess the limitations of currently available public datasets, and show what potential they do have to expand research into diversity-aware recommender systems.  Second, we analyse why and how European law and policy can be used to provide researchers with structural access to the data they need to develop diversity-aware news recommender systems.
\end{abstract}


\maketitle

%\section*{Impact Statement}
%The European Union has in recent years more actively stimulated technological innovation in the media, including through the Media Action Plan and the establishment of a Media Data Space. However, it is often still unclear what specific data is needed to develop responsible technologies for the media. This article focuses on news recommender systems, a key technology used by media organisations to determine what news their readers see. It shows the data needed to train diverse recommender systems is largely unavailable in existing datasets published by private media and technology companies. It further argues European law and policy that expands access to data is unsuitable to make the data needed to design diverse news recommender systems accessible. The article closes by providing a framework showing what data would need to be included in a dataset to enable the development of diverse recommender systems, and arguing policymakers should use the public service media to ensure the publication of such a dataset.  

%\localtableofcontents

\section{Introduction}
\label{sec:intro}
News recommender systems play an important role in determining what news (if any) individuals see online by ordering information from most to least relevant \cite{Helberger2016Policy,Loecherbach605Unified,2015Recommender}. Researchers have extensively criticised the common industry practice of measuring relevance through clicks, arguing this leads media organisations to recommend news articles based on what is most engaging rather than based on the editorial values that should guide the way the way the media informs the public.~\cite{Bernstein2020Diversity,Dacrema2019Are}. Over the past decade, researchers have therefore proposed alternative approaches to news recommendation that would support the media's democratic function of informing the public, particularly by allowing media organisations to make more diverse recommendations that expose each reader to perspectives and topics that are new to them. Researchers  have analysed extensively how diversity-aware recommender systems should be conceptualised \cite{Bernstein2020Diversity,Helberger2019On,Sax2022Algorithmic,Vermeulen2022Access,Vermeulen2022To}, and what technical tools are needed to measure and embed diversity in recommender systems \cite{Heitz2022Benefits,Lu2020Beyond,Vargas2011Rank,Vrijenhoek2022RADio,Bauer2024Where}.

However, a significant gap remains between the diversity-aware news recommender systems proposed in normative literature, and the kinds that have been realized in technical literature. For example, scholars in law and journalism studies call for recommender systems that allow readers to become experts in their chosen field by showing them new perspectives on a few topics that interest them, or that foster tolerance by showing readers political viewpoints with which they disagree but which do not lead them to see other societal groups as enemies \cite{Helberger2019On,Sax2022Algorithmic,Lin2022One}.  In contrast, technical literature most often implements diversity as an intra-list similarity function, to check whether the items in the recommendation are not too similar to each other \cite{castells2021novelty,raza2022news}. While work that designs news recommender systems based on more normative conceptualisations of diversity exists, it remains highly limited and regularly emphasises it is not yet suitable for deployment. Much work has yet to be done before industry-level recommender systems can truly account for normative interpretations of diversity~\cite{Vrijenhoek2022RADio}. 

In this paper, we argue that the public datasets on which researchers rely to create and evaluate news recommender systems are a key factor limiting the development of diversity-aware recommender systems. We show that while these datasets do have some untapped potential to further research into diversity-aware recommender systems, they lack much of the data needed to create the kinds of recommender systems proposed based on normative theory. Instead, they primarily facilitate the development of recommender systems that optimise for readers' short-term preferences. Further, we argue that this constraint cannot be fully addressed with the publication of another single dataset, as the data needed for diverse recommendations varies significantly across time and regions. We therefore also explore the European law and policy tools available to secure structural access to the data needed to develop diversity-aware news recommender systems. 

With this, we aim to answer the following research question: how do public datasets constrain the development of diversity-aware news recommender systems, and how can European law and policy be used to overcome this constraint? We focus on European law because of its recent advances in data access law and policy, its role for active governmental intervention to support media pluralism, and its wide geographic scope (allowing it to, potentially, support the creation of public datasets in multiple countries). We include the four most-used  public datasets in our analysis, but focus in particular on the Microsoft News Dataset (MIND). Due to its size, its publication in 2022 was a significant development for news recommendation research, and our research indicates that it is by far the most used dataset to develop new recommender systems. 

Section 2 analyses how public datasets are used to develop news recommender systems, and how they can impact the recommender systems used by media organisations by influencing the research that develops and evaluates the performance of new recommender systems.  Section 3 shows what data would have to be included in a dataset to develop the kinds of diversity-aware recommender systems imagined in legal and journalism studies literature. Section 4 evaluates how the data included in the public datasets that are currently most used in news recommendation research limits the development of diversity-aware recommender systems. Section 5 analyses why European law and policy should play a role to address the lack of datasets needed to build better news recommender systems, and how existing European laws and policies intended to expand access to data can do so. Throughout, we aim to make the paper accessible for  computer science, legal, and journalism scholars researching how more diverse news recommender systems can be realised. The analysis may therefore at times be overly simplistic for one specific audience. 
\section{From datasets to news recommendations}
\label{sec:datasets_to_recommendations}
Training a recommender system involves optimising it to predict a quantitatively measurable target; most often, whether a user will click on an item or not~\cite{zangerle2022evaluating}. A recommender system is trained to make these predictions by identifying patterns between item characteristics and user characteristics that are associated with the click. These patterns are subtle. For example, a click on an article about the Olympic Games may indicate the user is interested in the Olympic Games, the politics surrounding it, or a specific team, sport, or athlete. The broader the scope of content a system needs to cover, the harder it is to find the patterns that predict user behaviour. Furthermore, the way readers interact with a system is highly connected to how content is shown to them, meaning that slight changes in the user interface can lead to very different engagement patterns \cite{beel2021unreasonable}. To capture these nuances, recommender systems are trained on large datasets.  

Data at this scale is hard to come by: one needs to have a recommender system running in order to generate a dataset needed to train recommender systems. Media organisations resolve this chicken-and-egg situation by using simpler recommender systems to generate their own datasets, rather than relying on pulic datasets directly. This is due to the aforementioned impact that the user interface has on the way users interact with a recommender system: using a public dataset such as MIND (which was generated with \url{microsoftnews.msn.com}) to train a recommender system that will run on the BBC will mean the BBC's recommender system is optimised to identify patterns that differ from those in the context in which it is deployed. 

Researchers, however, rarely have direct access to the datasets needed to develop news recommender systems. Even if they do, relying on a proprietary dataset limits the reproducibility of their results \cite{Dacrema2019Are}. As a result, researchers and practitioners that create and test new recommender system algorithms generally rely on publicly available datasets. The research they produce on how recommender systems can be built, and which approaches are most successful, is in turn a valuable resource for media organisations, as it saves them from developing their recommender systems from scratch  \cite{Dodds2024Collaborative,Zamith2023Open}. 

One concrete example of how this works in practice can be seen in work by Einarsson, Helles \& Lomborg, which studied the effect of news recommender systems deployed by Ekstra Bladet during the 2022 Danish election \cite{Einarsson2024Algorithmic}. Ekstra Bladet had two recommender systems in production, both of which were trained on Ekstra Bladet's own data. However, the algorithms Ekstra Bladet used to train its recommender systems were developed and evaluated in the research literature, using public datasets (NRMS and Matrix Factorization \cite{Raza2021News,Wu2019Neural}. By analogy, Ekstra Bladet used a recipe (researchers' recommender system algorithm) that was developed and evaluated using one set of ingredients (public datasets) to create a specific dish (Ekstra Bladet's recommender system) using their own ingredients (Ekstra Bladet's  own dataset). 

In short, the research media organisations rely on to determine how to design recommender systems and which algorithms are most successful heavily relies on public datasets. Public datasets can shape this research in two main ways. Firstly, the data included in a dataset constrains what recommender systems developed with the dataset can be designed to do. ‘Simply' training a recommender system that predicts what news users will click will already require a large dataset of news articles and users' interactions with those articles from which these patterns can be derived. Developing a recommender system that balances user preferences with normative values such as diversity requires a richer dataset, which also includes data necessary to assess how the recommendation relates to those values \cite{Chin2022datasets,Stray2023Editorial}. For example, training a recommender to expose users to a diverse set of viewpoints requires a training dataset that contains data about the viewpoints expressed in its articles, or an effective proxy for this data. 

Second, a dataset can become a benchmark. Indeed, MIND was created with the explicit purpose ``to serve as a benchmark dataset for news recommendation and to facilitate the research in news recommendation" \cite{Microsoft}. Benchmark datasets fulfil an important function in research: they serve as a common reference point by allowing the performance of new recommendation algorithms to be compared to that of existing algorithms tested on the same dataset. However, they also make it more likely that research is done on the data structures present in the benchmark dataset. A good example of the practical implications of this dynamic is the MovieLens dataset, published in 2005 and still a widely used dataset for recommender system research. MovieLens contained information on users rating a movie on a scale of 1 to 5, which meant that consequently, a lot of research was carried out on predicting scores on an ordinal scale \cite{Harper2015MovieLens}. 
\section{Conceptualising diversity in news recommender systems}
\label{sec:conceptualising_diversity}
\subsection{Data as a constraint on diversity-aware news recommender systems}

Scholars from law and journalism studies have proposed many different ways in which recommendations can be made more diverse, including not only accounting for the viewpoints but also the topics, styles, and formats of the recommended content, as well as the characteristics of the users to whom content is recommended. Indeed, there arguably is not one right way to approach diversity. As the need for diverse news is often ultimately grounded in democratic theory, the type of diversity a recommender system should promote depends on the kind of democratic system (e.g., liberal, deliberative, or agonistic democracy) it is used to support \cite{Helberger2019On}. Our goal here is not to advocate for one specific approach to diversity. Nor do we argue a dataset should necessarily be suitable to realise all types of diversity we will outline below. For example, the need for diversity may be outweighed by the privacy concerns of collecting detailed data on the news consumption of marginalised groups. At this stage, we only aim to identify the characteristics of recommendations that are important from normative perspectives on diversity, so we can evaluate which aspects of diversity can and cannot be incorporated in news recommender system research with the data included in currently popular public datasets. 

\subsection{What data is needed to make diverse recommendations?}
\label{sec:what_data_is_needed}

The diversity of the topics of recommended articles plays a dominant role in the literature on diversity-aware news recommender systems. Many authors highlight that by accounting for what users have seen in the past, recommender systems could expose them to topics that are new to them, and thereby broaden their horizons or alert them to topics affecting other societal groups \cite{Harambam2018Democratizing,Sullivan2019Reading,Vermeulen2022To}. Alternatively, recommender systems can show a user extensive information on a few topics to create deeply informed expert citizens, or show individuals news on the topics they prefer to engage with to support their autonomy \cite{Helberger2019On, Hyzen2025Epistemic}. Between these extremes sit views that argue for a more mixed approach. A mix of entertaining and hard news could make diversity-aware recommender systems pleasurable to use, while a mix of political and non-political content could give individuals a fuller picture of society \cite{Bodo2019Selling,Harambam2018Democratizing,Helberger2019On}.

Viewpoint diversity is often addressed in conjunction with topic diversity, and concerns the diversity of perspectives on a particular topic to which users should be exposed \cite{Hada2022Beyond}. In general, authors advocate for showing users a wide diversity of perspectives within the same topic, either to better inform them, or increase tolerance and social cohesion \cite{Bernstein2020Diversity,Helberger2019On,Makhortykh2020Personalizing,Vermeulen2022To}. The different viewpoints to which users should be exposed are usually ideological in nature; several authors explore how users should be exposed to perspectives from across the political spectrum \cite{Bernstein2020Diversity,Heitz2022Benefits,Vermeulen2022To} or to the perspectives of other societal (e.g., ethnic, linguistic, or marginalised) groups to increase tolerance \cite{Bastian2019News,Makhortykh2020Personalizing,Sax2022Algorithmic}. An important but rarely explicit assumption in this context, is that the diversity of perspectives to which a recommender system should expose users may differ depending on the political perspectives or societal groups that exist in the society in which the system is deployed \cite{Vrijenhoek2024Diversity}. 

The style of news plays an important role in discussions on the type of public debate diversity-aware recommenders should foster. Deliberative approaches that see media as creating space for an informed and rational debate imply the recommended content should be impartial and promote active discourse \cite{Helberger2019On}. Recommender systems that seek to foster tolerance may recommend positive news about ideological opponents, or respectful discussions between opponents \cite{Harambam2018Democratizing}. Finally, from an agonistic perspective the goal of a recommender should be to facilitate conflict while ensuring the different sides do not perceive each other as enemies \cite{Sax2022Algorithmic}. 

The format in which news is recommended matters to recommender systems that respond to users' consumption preferences. If a recommender system is used to serve a wide cross-section of society, it is important to ensure news is recommended in a format (such as video, audio, or text) with which different users can engage \cite{Council2018Recommendation,Helberger2019On,Stray2023Editorial}. The need for different formats is also implied in some of the approaches to diversity mentioned above. For example, a recommender system that aims to enable users to become experts in a specific topic requires larger background articles and deep dives for users that have a lot of pre-existing knowledge on the topic, and simple explainers for users that do not. 

Finally, few authors pay explicit attention to the characteristics of the users to whom content is recommended. One exception is Vermeulen, who emphasises the need to balance diverse exposure with user choice \cite{Vermeulen2022To}. Additionally, detailed information about the user is required to provide the diverse recommendations described above. For example, a recommender system must be able to assess what news a user knows already in order to give them a new perspective or more deeply inform them. From a fundamental rights perspective, moreover, the need for diversity is also grounded in the right to receive information, and more specifically the need to ensure that ``everyone has access to a diverse range of journalistic content” \cite{Eskens2017Challenged,Council2022Recommendation}. In that context the Council of Europe has for example referred to the need to ensure diverse news can be accessed despite socio-economic barriers and income level and that minorities, disabled persons, and disadvantaged and local communities are able to access news \cite{Council2018Recommendation,Council2022Recommendation}. Similarly, writing on recommender systems' impact on epistemic welfare, \cite{Hyzen2025Epistemic} emphasise the importance of ensuring a recommender system is able to ``ensure that many and diverse users can encounter and engage with epistemically valuable content". Overall, if the goal of a recommender system is to meet the information needs of all members of society, it is also important for the recommender system to be able to effectively recommend news to different societal groups, even if they have different consumption patterns. 

\subsection{Subconclusion}
Many of the features that would be needed to make diverse recommendations are hard to identify, and require additional data processing \cite{Vrijenhoek2021Recommenders,Vrijenhoek2022RADio}. Think, for example, of identifying the opinions reflected in an article text so a recommender system can show a user different perspectives on the same topic. There are extensive lines of research into identifying different viewpoints algorithmically, usually referred to as opinion mining or viewpoint detection. However, neither currently have off-the-shelf solutions \cite{Reuver2021Are}. Alternatively, researchers or journalists could manually annotate the viewpoints included in an article \cite{Mascarell2021Stance}. However, manual annotation at the scale needed for training data for recommender systems is a time-consuming and costly endeavour. This problem is exacerbated by the conceptual unclarity of diversity, as labelling by non-experts is likely to yield inconsistent results. Despite, and arguably especially given the difficulty of identifying the features relevant to diversity, it is important that these aspects are included in public datasets, if they are to facilitate research developing diversity-aware recommender systems.
\section{Data as a constraint on diversity-aware news recommender systems}
\label{sec:data_constraint}
In this section we analyse which datasets are used in recent research developing and testing news recommendation models, the characteristics of those datasets, and how they can(not) facilitate the different aspects of diversity outlined on page \pageref{sec:conceptualising_diversity}. 

\begin{figure*}[h]
    \centering
    \includegraphics[width=0.75\linewidth]{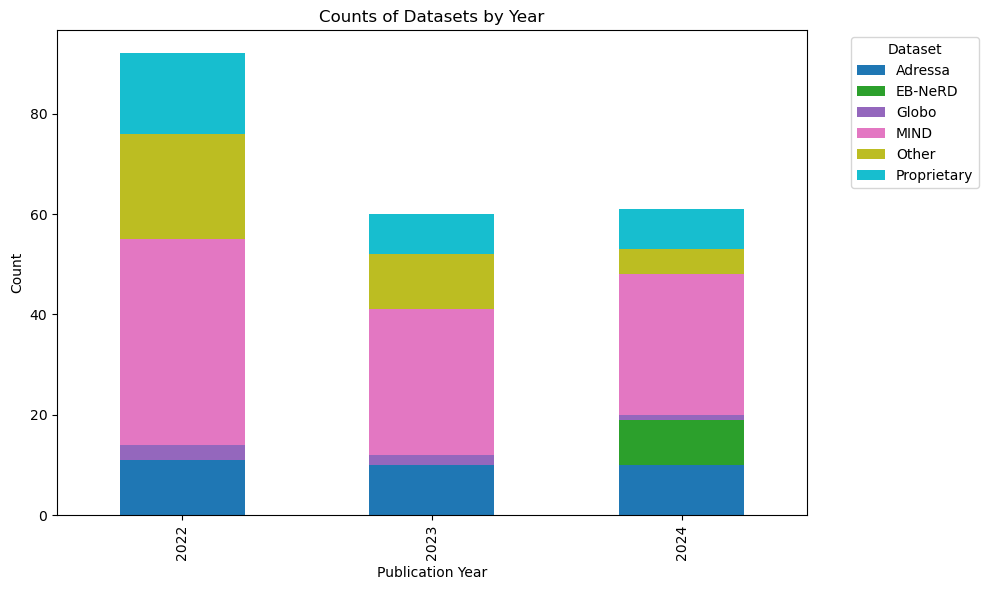}
    \caption{Counts of the datasets used in papers on developing news recommender systems between January 1st, 2022 and December 31, 2024. Excludes datasets cited 5 times or fewer.}
    \label{fig:datasets}
\end{figure*}

\subsection{Datasets used in current news recommendation research}
To assess which public datasets are used for news recommendation research, we queried the Scopus search engine for papers with the term ``news recommend*" in the title, abstract or keywords, that were published between January 2022 and December 2024. This returned 383 results. We filtered out conference proceedings, papers that only mention news recommendation as a potential field of application, that are behind a paywall, or not written in English. This left us with 314 papers. Out of these 314 papers, 46 are written from a social science perspective, whereas the vast majority (268) is technical. The majority of technical papers (171) implemented and tested a news recommendation model. 

Figure \ref{fig:datasets} provides an overview of the datasets used in these papers. It lists all datasets that are cited more than 5 times, and groups the others into the `Other' and `Proprietary' categories. The `Other' category (35 papers) mostly consists of datasets such as MovieLens, which is a standard benchmark dataset for (non-news specific) recommendation tasks, and task-specific datasets such as FakeNewsNet for fake news detection. In addition, the `Proprietary' category (32 papers) covers datasets that are not shared with the academic community. Such datasets are always used for online experiments that test recommender systems on real users, but without publication of the data these studies are not reproducible.

There are four news-specific public datasets that are frequently used to implement and test news recommendation models: MIND \cite{Wu2020MIND}(used 98 times), Adressa \cite{Gulla2017Adressa}(used 31 times), EB-NeRD \cite{Kruse2024EB}(published for the 2024 RecSys challenge, used 9 times), and Globo\cite{Souza2018News} (used 6 times). All four datasets contain (the IDs of) the articles people clicked during a certain time period on a news website: the fully algorithmic aggregator MSN News for MIND, Norwegian Adresseavisen for Adressa, the Brazilian G1 news portal for Globo, and Danish tabloid Ekstra Bladet for EB-NeRD. Table \ref{tab:datasets} provides an overview of the data contained in each of the datasets.

MIND is the most prominent dataset, accounting for 57\% of all papers published on developing news recommendation models. The publication of MIND in 2020 was a significant development for news recommender system research. While at first glance its scale seems comparable to that of Adressa, the interactions in MIND take place in a much shorter time period. This means that there are more interactions per published news article. The articles in MIND are written in English, which increases result interpretability for researchers. Most importantly, and contrary to Adressa and Globo, MIND also contains an `impression log' with articles that users saw but did not click. In addition to this raw data, Microsoft published a package of news recommender algorithms that could be trained on the dataset through open source code. Microsoft also launched a competition, challenging developers to create algorithms that would outperform their own. Performance was determined by the recommender system's ability to correctly rank a list of candidate items based on the likelihood a user clicks on them.

% Please add the following required packages to your document preamble:
% \usepackage{graphicx}
\begin{table*}[]
\footnotesize
\begin{tabularx}{\linewidth}{ X X X X X X }
\hline
 & MIND & Adressa & Globo & EB-NeRD \\ \hline
\#interactions & 15 million & 23 million & 3 million & 37 million \\ \hline
\#users & 1 million & 3 million & 314.000 & 1 million \\ \hline
\#articles & 160.000 & 48.000 & 46.000 & 125.000 \\ \hline
time period covered & Oct 12 - Nov 22, 2019 & Jan 1 - Mar 31, 2017 & Oct 1 - Oct 16, 2017 & Apr 27 - June 8, 2023 \\ \hline
language & English & Norwegian & Portuguese & Danish \\ \hline
organization type & aggregator & local news & national news & national tabloid \\ \hline
article metadata & category, subCategory, title, abstract, URL,  named entities in title, named entities in abstract, embeddings & title, category, URL, word count, time published, keywords, author, entities & category, time published, publisher, word count, embeddings & title, subtitle, body, category, subcategory, time published, premium or open, IDs of corresponding images, article type, topic, views, reads, sentiment, entities, embeddings \\ \hline
click metadata & user ID, timestamp, clicked articles, non-clicked articles & timestamp, session start/stop, read time, referrer & user ID, article ID, timestamp, session, referrer & user ID, article ID, session ID, articles viewed, articles clicked, time stamp, read time, scroll percentage, device type \\ \hline
user metadata & history & city, region, country, os, device & region, country, os, device & history, subscription, gender, post code, age \\ \hline
\end{tabularx}%
\caption{The context and data included in the four primary datasets used for training news recommender systems. Most of the users on news websites are not logged in - unless IP address is considered it is difficult to determine the number of unique users a dataset encompasses. The URLs to the full articles in MIND and Adressa are expired (\protect\url{https://github.com/msnews/msnews.github.io/issues/22}).}
\label{tab:datasets}
\end{table*}

\subsection{The constraints of current datasets from a diversity perspective}
The datasets described above are important resources in news recommender system research and development - it would be very challenging for researchers to develop (reproducible) recommendation models without them. However, as we argue below, the characteristics of these datasets (specifically the kind of metadata, content and audiences they contain) pose significant limitations to the development of diversity-aware recommender systems, and make it comparatively easier to develop recommender systems optimised for engagement. Our analysis includes all datasets but focuses in more detail on MIND, as our research indicates this is the most frequently used public dataset in news recommender system research.

\subsubsection{Metadata}
Most datasets contain only very limited information about the articles that have been recommended. None of the datasets contain readily available information on the styles and formats of the articles, or the perspectives expressed in them. This makes it considerably more difficult to use these datasets to develop recommender systems that incorporate these aspects of diversity. As mentioned in the previous section, it is possible to obtain information on for example the viewpoint expressed in an article by mining the relevant content characteristics from the text of article. However, this process (already difficult in itself) is complicated by the fact that, with the exception of EB-NeRD, the datasets do not contain the full text of the articles. Because of licensing issues, Globo has not published any additional information about their articles; content is only represented through an ID, category code (e.g,.`614') and embeddings created based on all the texts. MIND and Adressa only contain an URL, not the actual content of news articles. Before researchers begin the difficult work of mining characteristics from articles' content, they must therefore verify whether they are legally allowed to scrape the content, build the tools to do so, and ascertain which URLs still function (which, in May 2025, is not the case in either of the datasets). 

Adressa, EB-NeRD and MIND all do contain metadata on an article's topic, category, and title; Adressa also has information on the author of an article. This information could be used to develop recommender systems that recommend diverse topics or categories of articles to users. However, the generic nature of the metadata that is available on the articles' topics challenges the development of such recommender systems. A recommender system can only identify patterns in data that is available; as such, the granularity with which article categories are chosen significantly influences the types of patterns that can be identified. For example, all political news in MIND is put under the umbrella category `politics'. As such, a recommender system developed with MIND could not determine a preference for the Democratic over the Republican Party, let alone a preference in non-American politics, unless other metadata such as the title contains sufficient data to infer such specific information. 

\subsubsection{Content}
The content in the datasets is another significant limitation on the datasets' suitability for the development of diversity-aware recommender systems. To train a recommender system to recommend diverse topics or perspectives on matters of public interest (whether to promote social cohesion, create expert citizens, or another objective of diversity-aware news recommenders), the recommended articles must actually include diverse content and perspectives. However, only 30\% of the MIND's dataset consists of articles categorised as news, with the rest fitting into categories such as `lifestyle', `sports', and `food' \cite{Vrijenhoek2023Do}. Similar distributions can be found in EB-NeRD, where only 28.000 of the 125.000 articles belong to the `news' category. Adressa contains roughly 60\% news, but its categories are mostly focused on the geographic area the article covers, rather than its content (`nyheter|nordtrondelag'). Furthermore, the subcategories MIND uses for news indicate it is focused on the US context, with `newsus' representing 47\% of all news items, `newspolitics' 17\%, and `newsworld' 8\% \cite{Vrijenhoek2023Do}. 

This is potentially problematic; while the datasets contain large amounts of articles, they contain relatively few news articles, which are likely to include the kinds of diverse ideological perspectives and address the kinds of public interest issues that the diversity-aware recommender systems proposed in normative literature take into account. This issue is exacerbated by the popularity of MIND and its focus on US news, as a recommender system trained to recommend diverse political perspectives and topics in the US two-party system would be optimised to identify consumption patterns that differ from countries with different political and societal groups \cite{Shivaram2022Reducing}. This would in turn challenge the ability of recommender systems to facilitate the access of individuals outside the US to content that reflects the diversity of political outlooks or societal groups in the society in which they live. This issue can be partially addressed by use of the other datasets, which reflect the Brazilian, Norwegian and Danish political systems. However, the popularity of MIND indicates this solution has not yet been widely adopted. 

Finally, it should be noted that EB-NeRD has some unique potential for the development of style- and format-aware recommender systems. Unlike the other datasets it contains additional information like content type (video, gallery, etc.), accompanying images, and estimated sentiment score. It could therefore be used to research recommender systems that respond to the format consumption preferences of different users (e.g., deep-dives or simple explainers) and affective styles (e.g., deliberative discussion or humanisation of political opponents). Conversely, the lack of such data in MIND, as well as the fact that MIND only contains text articles, make it more difficult to use this dataset to take the style and format of news into account when developing diversity-aware news recommenders. 

\subsubsection{Audience}
There are distinct differences in the information each of the datasets includes about its users. The only information available on MIND's users is when they accessed the system, which items they have interacted with in the past, which items were presented to them in a specific session, and on which items they have clicked. This is commendable from a privacy perspective. However, it makes it impossible to assess for which audiences recommender systems trained on MIND are optimised. Different societal groups (e.g., age, ethnic, or socio-economic background) are likely to have different consumption patterns~\cite{vandenbroucke2025not}. If a particular user's or group's consumption pattern falls outside the patterns for which a recommender system is optimised, the recommender system will be less able to determine what news they should be recommended. It is likely that MSN users are not fully representative of the different audiences for whom recommender systems are developed, if only because of MIND's focus on English-language US news. However, as MIND does not contain even aggregate data on the characteristics of its users, it is impossible to know exactly in what ways and for which groups MIND's audience is unrepresentative. The other datasets do contain information on the geographic location a click happened: Globo provides the region, Adressa the city, and EB-NeRD the postcode. This data makes it possible to assess how a recommender system performs for local communities in each country. However, it is not possible to assess how a recommender system performs for different socio-economic, disadvantaged, or minority groups. EB-NeRD is a partial exception to this, as postcode could be combined with other data to infer (for example) socio-economic status and the dataset sometimes contains information on users' gender (7\%) and age group (3\%). 

The limited number of days in each of the datasets poses another limitation. MIND and EB-NeRD only contain interactions of a period of 6 weeks, versus two weeks for Globo and three months for Adressa. Moreover, as users of news websites are often not logged in (6.7\% in EB-NeRD), it is difficult or impossible to track changing behavior over time. For example, in MIND only 20\% of visits in the dataset are from users that have accessed the system five or more times, which can be explained by the fact that user IDs reset after 24 hours. As such, little of the data that is necessary to reliably observe long term changes in people's reading behavior is available. This limits research on diversity-aware recommender systems that aim to foster long term changes in the audience, for example to promote tolerance, create expert citizens, or stimulate self-development. 

Furthermore, all datasets encompass a time period where, electorally, nothing significant happened. It is therefore still impossible to gauge how a model would behave during democratically important moments such as elections. Most of the time, this is a deliberate design choice. On the challenge website EB explicitly states that "[t]his timeframe was selected to avoid major events, e.g., holidays or elections, that could trigger atypical behavior at Ekstra Bladet." However Einarsson, Helles \& Lomborg, in a prior direct collaboration project with Ekstra Bladet, noted that during the 2019 Danish election the use of news recommender system increased the amount of soft news readers consumed \cite{Einarsson2024Algorithmic}. If diverse recommendation algorithms are to be researched and developed, they also (perhaps even especially) need to be robust towards exceptional behavior patterns.

\subsection{Implications for research} 	

Taken together, the data included in the currently most-used datasets are most suitable to develop recommender systems that prioritise news based on users' immediate, short term content preferences. Much of the data needed to develop the kinds of diversity-aware recommender systems proposed in legal and journalism studies literature based on democratic principles is lacking from public datasets. In particular, the recommendation of diverse perspectives, formats, and styles is highly challenging with current datasets, as the relevant content characteristics are either not included as metadata or lacking from the recommended articles themselves. The development of recommender systems that aim to foster long term changes (for example by promoting tolerance or deeply informing individuals about specific topics they select) or are robust enough to perform well during different societally significant events (such as elections or wars) is similarly challenging, due to the short time period covered by the datasets. To a lesser extent, optimising recommender systems to perform well for different societal groups is difficult with current public datasets, as data on the individuals to whom news is recommended is often lacking from the datasets. 

The lack of data relevant to diverse recommendations also shapes the way datasets can be used to evaluate the performance of recommender systems. This is particularly important for benchmark datasets that are to serve as a common reference point against which the performance of all recommender systems is measured. In the context of MIND, the success of recommender system algorithms is judged by how well they predict the way 1 million MSN News users in 2019 interacted with the English-language mix of hard, entertainment, and sports news on that site. This is not a neutral measure of success. Because no data is available on the characteristics of these MSN users, it is not clear how they are (un)representative of other groups. However, for example, a recommendation algorithm that is perfectly able to inform expert and non-expert users in the Netherlands by providing them news at the level of detail they are able to understand may not perform well on MIND, due to the limited diversity of its content and likely focus on a US audience.

The datasets' suitability for creating recommender systems that optimise for engagement is exacerbated by the competitions set out by the companies that have published them.  Competitions are a common way for companies to increase researchers' use of their dataset, and allow them to set out the specific performance indicator they would like researchers to optimise for \cite{Luitse2024Platform}. In the case of MIND, Microsoft set out a competition tasking researchers to create a recommender system that ranks ``articles according to the personal interest of [the] user" \cite{Qiao2020MIND}. The 2024 RecSys challenge, which focused on news recommendation and relied on EB-NeRD, did aim to address ``both the technical and normative challenges inherent in the design of effective and responsible recommender systems". However, the main evaluation metric used to determine the winner of the competition focused on engagement: the odds that a randomly chosen clicked article is ranked higher than a randomly chosen non-clicked article (referred to in computer science literature as the `area under the curve' \cite{Wu2019Neural}). While participants were encouraged to also include beyond-accuracy objectives in their challenge submission, participants were not evaluated on these metrics, and challenge submissions exclusively maximised accuracy. One of the challenge participants (\cite{heitz2024recommendations}) noted that, through the design of the challenge the potential to optimise on diversity was limited, and that therefore the challenge was ``a missed opportunity to move beyond `leaderboard-chasing culture'".
\section{Ensuring access to training data: the role of European law and policy}
\label{sec:role_law}
\subsection{The legal case for public support for public datasets}

If one were to construct a dataset that is better able to enable the development of diverse news recommender systems proposed in legal and journalism studies literature based on democratic principles, it would include 1) a focus on news relevant to public debate, 2) granular metadata on the content characteristics (topic, viewpoint, format, and style) that a diversity-aware recommender may need to take into account, 3) a large number of frequently returning users over a longer period of time, and 4) with sufficient metadata on these users and the societal groups they represent. Table 1 provides a more complete overview of the data needed to develop diverse recommender systems, based on the analysis in sections 3 and 4.

% Please add the following required packages to your document preamble:
% \usepackage{graphicx}
% \usepackage[normalem]{ulem}
% \useunder{\uline}{\ul}{}
\begin{table}[]
\centering
%\resizebox{\textwidth}{!}{%
%\begin{tabular}{c|l}
%\begin{tabularx}{\linewidth}{c|p{0.7\linewidth}}
\begin{tabular}{c|p{0.75\linewidth}}
\textbf{Aspect of diversity} & \textbf{Data to be included} \\ \hline
Topics & News on the political topics about which citizens need to be informed in their democratic society; topics affecting other societal groups; topics (e.g., entertainment, fashion, climate) with which individuals want to engage. \\
Perspectives & The range of ideological and political perspectives in society; perspectives of societal groups (e.g., ethnic, linguistic, or marginalised) groups. \\
Style & Impartial and promoting active discourse; Positive and respectful about societal opponents; critical about opponents while not framing them as enemies. \\
Format & Video/audio/text to meet the consumption preferences of different news users; deep dives and simple explainers depending on users’ pre-existing knowledge about a topic. \\
Users & Different groups (e.g., socio-economic groups, minorities, disabled persons, and disadvantaged and local communities) that need to be able to be informed in democratic society; data on user characteristics relevant to the above, such as the formats they prefer to consume, political leaning, topics they have engaged with.
\end{tabular}%
\caption{Overview of data to be included in datasets to facilitate the development of diversity-aware recommender systems proposed in legal and journalism studies literature}
\label{tab:placeholder}
\end{table}

It is unlikely any single dataset will contain all of this data. As noted in section 3, different data is required depending on the specific kind of diversity a recommender system is expected to promote. More importantly, however, the the viewpoints and topics present in the public debate, as well as the societal groups whose perspectives are recommended and for whom recommender systems have to be optimised are different in each society and at different times. While the publication of another dataset with the characteristics above would already be a significant improvement, it would only facilitate the development of diversity-aware recommender systems that optimise their recommendations for the specific society and the specific time in which the dataset was created. To enable the development of recommender systems that work for multiple different societies, multiple datasets are needed. 

There is little incentive for any particular media organisation to invest in the creation of a high quality dataset others could use to research diversity-aware recommender systems, much less the creation of multiple datasets that reflect the consumption patterns of different societies at different times. Nor is facilitating the development of diversity-aware recommender systems arguably the societal role of a commercial media or technology organisation. However, states do have an interest in ensuring that the different media organisations in their society can provide diverse perspectives. Under European fundamental rights law, this is moreover a legal obligation. Article 10 of the European Convention on Human Rights has long imposed a positive obligation on states to be the “ultimate guarantor” of diversity, meaning states must not only refrain from censorship, but also actively ensure that media organisations can provide (and the public can receive) diverse perspectives representing different societal groups \cite{2009Manole,2022NIT,Spano2023concept}. 

States have traditionally fulfilled this obligation by ensuring the existence of different media organisations through subsidies or media concentration laws, or by creating public service media organisations. However, as scholars have argued extensively over the past decade, individuals' access to diverse information not only relies on the existence of different media organisations, but also on the design of technologies such as recommender systems that determine to what information individuals are exposed \cite{Brogi2021EU,Council2018Recommendation,Helberger2012Exposure}. So far, legal researchers and policymakers have primarily addressed the media's use of technology by emphasising the media's responsibility  to use technologies in line with editorial values, as well as the limits freedom of expression imposes on any regulation of the media's use of technology \cite{Council2023Guidelines,Helberger2020freedom,Ofcom2024Ofcoms,Vermeulen2022Access}. However, both to ensure media organisations can actually live up to these responsibilities and to fulfill their own obligation to guarantee diversity in the media system, it is also important to consider how states can create the conditions media organisations need to use technologies like recommender systems responsibly. In particular, as we have argued in this article, the existence of a high quality dataset is an important condition for organisations to ultimately be able to recommend diverse news to their audiences. 

Law and policy that supports the creation of public datasets also serves another purpose, namely to safeguard the media's independence from large technology companies. A large strand of journalism studies research has assessed how media organisations rely on large technology companies for technologies used to gather, produce, and distribute news, including news recommender systems \cite{Kristensen2023Infrastructure,Simon2024Artificial,Santos2023Google,Fanta2020Google,Papaevangelou2023Funding}, as well as the computing power, research, and indeed training data needed to develop new technologies \cite{Simon2022Uneasy,Simon2023Escape,Simon2024Artificial}. This reliance challenges media organisations' ability to independently determine how the technologies they use inside newsrooms serve editorial values. Instead, they increasingly adopt the logics of the large technology companies on which they rely \cite{Santos2023Google,Luitse2024AI,Simon2022Uneasy,Papaevangelou2023Funding,Drunen2021Editorial}.  As our research shows, over half the research into the development of new news recommender systems relies since 2022 relies on a dataset published by Microsoft (MIND), and the datasets currently available are primarily suitable for developing recommender systems that optimise for engagement. As such, public support for the creation of datasets that facilitate the creation of recommender systems that promote different conceptualisations of diversity would lessen MIND's influence on the way media organisations can recommend news to their audiences. 

\subsection{Access to public datasets through European data access law and policy}
EU law has significantly expanded its data access obligations in recent years. However, none of the new obligations ensure access to data that could be used to develop diversity-aware recommender systems. The Data Governance Act requires public sector bodies to facilitate the re-use of data, but article 3(2) exempts public service broadcasters and cultural institutions from this obligation. The Data Act's data access obligations are limited to the Internet of Things (Chapter II) or public authorities (Chapter V). Finally, article 40 of the Digital Services Act does grant researchers access to data necessary to assess the effects of large platforms' recommender systems on diversity. However, this data access right is limited to the data that is necessary to understand platforms' impact and risk mitigation measures. While it could be used to evaluate the diversity of the datasets used to train platforms' recommender systems along the criteria laid out on page \pageref{sec:what_data_is_needed}, it therefore likely does not allow researchers to develop alternative recommender systems.

EU law and policy is also beginning to facilitate voluntary data sharing. Data spaces, which allow organisations to share data in accordance with common standards, play a particularly important role in the EU's push to strengthen digital media and the European technology industry more broadly  \cite{Commission2020Europes,Commission2022Commission}. The Commission has in 2022 set out a €8.000.000 grant for the development of a ‘media data space' that would provide organisations with content, metadata, and audience data to develop “data services matching European values, in particular ethics, equality and diversity”, according \cite{Commission2022Digital}. A pilot project and the European Broadcasting Union (an organisation representing Europe's public service media) both foresee that this media data space could provide access to data necessary to develop recommender systems \cite{Commission2022Pilot,EBU2022Dataspace,ECdatastories}. 

In large part, however, the focus on data spaces in European (media) data policy aims to address a different issue than the one posed by the lack of data needed to develop diversity-aware recommender systems. Data spaces are useful to increase access to data currently held by separate organisations. The same is true of other mechanisms in the Data and Data Governance Act facilitating data sharing, such as data intermediaries or data altruism. However, the differences in audience behaviour caused by (for example) different recommender systems' user interfaces, algorithms, and goals, make it highly challenging to identify meaningful patterns between recommendations and user behaviour in such an aggregated dataset. Data spaces may therefore be most suitable as an overarching framework through which datasets for recommender system development can be made available. For example, they can make it easier to test recommender systems on multiple datasets by setting common standards on data structure, and provide means for other actors to enrich datasets by adding metadata concerning, for example, viewpoint diversity \cite{Commission2022Pilot,EBU2022Dataspace,Matas2023Data}. They can also expand access to the data necessary to understand audience preferences and behaviours on a more general level \cite{ECdatastories}. However, all this would not address the core issue identified in this article, namely the lack of a dataset (or datasets) with which different types of recommender systems promoting different kinds of diversity can be developed. 

Though EU law does not currently ensure access to such a dataset, it also does not pre-empt either governmental or media initiatives to create it. In particular, public service media organisations are arguably well positioned to do so. Public service media organisations already have access to a diverse library of content that can be recommended, and several already operate recommender systems that can be used to generate datasets. Moreover,  in contrast to Microsoft and the commercial media organisations that have so published the datasets used in this article, public service media organisations have a specific societal role and sometimes even legal obligation to promote citizens' ability to access diverse information \cite{Donders2021Public,Vermeulen2022Access}. Public service media organisations have traditionally fulfilled this role by providing diverse content themselves. Increasingly, they have also experimented with using recommender systems to better serve their own audiences, and in this process been forced to navigate the difficulties of incorporating diversity in their recommender systems \cite{Hilden2022Public,Srensen2019Public}. 

However, what remains lacking, and what public service media organisations could begin to provide, is support for the more foundational research that enables media organisations throughout the media system to automate editorial decision-making in line with their editorial values. European law can only play a limited role in facilitating such support; the regulation of public service media remains primarily a national competence. However, organisations such as the European Broadcasting Union do already operate projects that aim to facilitate cooperation between public service media developing diverse recommender systems \cite{Hilden2022Public, PEACH}. Such approaches could be expanded to also generate publicly accessible datasets needed for researchers to create new kinds of diverse recommender systems. Alternatively, individual public service media organisations could publish datasets  on their own initiative. In both cases, the data outlined in table 2 may serve as a useful starting point for public service media organisations looking to publish the data needed to enable the creation of a wide variety of diversity-aware news recommender systems \cite{Stray2023Editorial}.
\section{Conclusion}
\label{sec:conclusion}
In this paper we have analysed how the lack of suitable public datasets constrains the development of diversity-aware recommender systems. We have analysed what data is needed to develop the kinds of diversity-aware recommender systems proposed in legal and journalism studies literature, and to what extent this data is available in the currently most used public datasets. We have argued current datasets are primarily useful to optimise for short-term engagement, and to a much lesser extent for the recommendation of diverse topics. The development of diverse  recommender systems that take the formats, styles, and viewpoints of articles or the characteristics of different audiences into account is particularly limited by current datasets. Furthermore, we argued why European law should address this limitation, how existing data access measures fail to do so, and how public service media playing a more active role in technological development could provide a way forward. 

In the short term, our analysis indicates there is potential to better use the currently available datasets to develop diversity-aware recommender systems. The majority of research in the past 3 years has relied on MIND; complementing the use of this dataset with EB-NeRD, Adressa, and Globo would allow research to optimise the performance of recommender systems not only for audiences that resemble 2019 US MSN news users, but also audiences from (specific regions in) Brazil, Denmark, and Norway. In addition, due to its inclusion of article text and wider variety of content, EB-NeRD partially allows for the development of diversity aware recommender systems that take the formats, styles, and viewpoints of articles into account by inferring the relevant characteristics from the text. However, the lack of metadata on (for example) the viewpoints represented in the articles still makes this challenging. Moreover, the limited spectrum of news content, formats, and timeframes in all datasets continue to limit the development of diversity aware recommender systems. 

In the long term legal and technical research should therefore also expand its focus from assessing how the media should use technologies such as recommender systems responsibly, to analysing the conditions that need to be in place for researchers to develop and media organisations to deploy technologies that align with editorial values. At present, the public datasets used to develop diversity-aware news recommender systems are supplied by commercial technology companies (Microsoft) or commercial media companies (Adresseavisen, Ekstra Bladet, and G1). Providing the datasets needed to develop the kinds of diversity-aware recommender systems proposed in normative theory is not these companies' societal role, nor do they (given their financial goals) have a strong incentive to do so. Instead, to enable media organisations to enact the responsible use they call for, as well as to live up to their own obligation under European fundamental rights law to safeguard a diverse media system, states should take a more active role in providing the data needed to develop diversity-aware news recommender systems. Collaboration between researchers and public service media organisations has an important role to play here, as these organisations have the diverse content and domain knowledge needed to generate the necessary data, as well as the societal role to promote access to diverse news.

\section*{Acknowledgments}
We would like to thank Natali Helberger and Laura Hollink for their feedback on earlier drafts of this article. All errors are ours. 

\paragraph{Funding Statement}
This research was supported by funding for the AI, Media and Democracy Lab (grant nr. NWA.1332.20.009) from the Dutch Research Council.

\paragraph{Data availability statement}
The four datasets studied in this article are available at https://msnews.github.io/ (MIND), https://recsys.eb.dk/dataset/ (EB-NeRD), https://www.kaggle.com/datasets/gspmoreira/news-portal-user-interactions-by-globocom (Globo), and https://reclab.idi.ntnu.no/dataset/ (Adressa). Restrictions apply to their  use.

\paragraph{Competing Interests}
Max van Drunen co-authored a chapter in a book projected funded by Microsoft while writing this article. Further information is available at https://www.ivir.nl/projects/news-and-media-law-in-europe/

\paragraph{Ethical Standards}
The research meets all ethical guidelines, including adherence to the legal requirements of the study country.

\bibliographystyle{plainnat}
\bibliography{References.bib}

%\begin{Backmatter}

%\renewcommand\bibpreamble{By default, this template uses \texttt{bibtex} and adopts the AMS referencing style. However, the journal you’re submitting to may require a different reference style; specify the journal you're using with the class' \texttt{journal} option --- see lines 1--19 of \emph{sample.tex} for a list of options and instructions for selecting the journal.}

% If using any of the following journal options:
%   wet, dap, dce, eds, prm, flw, jdm, psy, rsm
% then use the \printbibliography line instead of:
%\bibliography{example}
%\printbibliography

%\end{Backmatter}

\end{document}